\documentclass[a4paper]{article}

\usepackage{INTERSPEECH2022}
\usepackage{cite}

\title{PM-MMUT: Boosted Phone-Mask Data Augmentation using Multi-Modeling Unit Training for Phonetic-Reduction-Robust E2E Speech Recognition}
\name{Guodong Ma$^1$, Pengfei Hu$^2$, Nurmemet Yolwas$^{1,3}$, Shen Huang$^2$, Hao Huang$^{1,3*}$
\thanks{$^*$ corresponding author}}
\address{
  $^1$School of Information Science and Engineering, Xinjiang University, Urumqi, China \\
  $^2$Tencent Minority-Mandarin Translation, Beijing, China \\
  $^{3}$Xinjiang Key Laboratory of Multi-lingual Information Technology, Urumqi, China} 
\email{hwanghao@gmail.com}

\begin{document}

\maketitle
\begin{abstract}
Consonant and vowel reduction are often encountered in speech, which might cause performance degradation in automatic speech recognition (ASR). Our recently proposed learning strategy based on masking, Phone Masking Training (PMT), alleviates the impact of such phenomenon in Uyghur ASR. Although PMT achieves remarkably improvements, there still exists room for further gains due to the granularity mismatch between the masking unit of PMT (phoneme) and the modeling unit (word-piece). To boost the performance of PMT, we  propose multi-modeling unit training (MMUT) architecture fusion with PMT (PM-MMUT). The idea of MMUT framework is to split the Encoder into two parts including acoustic feature sequences to phoneme-level representation (AF-to-PLR) and phoneme-level representation to word-piece-level representation (PLR-to-WPLR). It allows AF-to-PLR to be optimized by an intermediate phoneme-based CTC loss to learn the rich phoneme-level context information brought by PMT. Experimental results on Uyghur ASR show that the proposed approaches outperform obviously the pure PMT.
 We also conduct experiments on  the 960-hour Librispeech benchmark using ESPnet1, which achieves about {10\%} relative WER reduction on all the test set without LM fusion comparing with the latest official ESPnet1 pre-trained model.
\end{abstract}

\section{Introduction}
\label{sec:intro}

Recently, end-to-end (E2E) automatic speech recognition (ASR) \cite{ctc-paper,RNNt-paper,LAS-paper,CTC_conformer,Speech-Transformer,Conformer} based neural networks have achieved large improvements. 
With the emergence of E2E ASR, researchers \cite{SpecAugment,SpecAugment_large_dataset,mixspeech_paper,semantic_paper,jicheng_paper,yizhou_paper,decouple_pronunciation,hybrid_modeling_units,augmentation_paper1,augmentation_paper2}  explore different E2E ASR scenarios and partly focus on the data augmentation and training strategy due to the nature of data hungry and easy over-fitting. 

Consonant and vowel reduction are often encountered in speech, especially in spontaneous audio, which might cause performance degradation in ASR. This issue is denoted by phonetic reduction (PR). However, few people pay attention to it.
To our best knowledge, perhaps our PMT \cite{Phone_mask} is the first to alleviate the impact of PR on the  E2E ASR system by randomly masking off a certain percentage features of phones and filling with the average value of word where the phone is located during model training, thereby enhancing the robustness of the model to PR. Although PMT mitigates the impact of PR on E2E ASR to some certain extent, we find that there are some disadvantages of the proposed PMT system. The general modeling unit is based on word-piece \cite{Uygur-paper1,Phone_mask,recent-conformer,english_word_piece1,english_word_piece2,english_word_piece3,english_word_piece4,english_word_piece5}. 
Therefore, it will cause granularity mismatch between the masking unit of PMT (phoneme) and  the modeling unit (word-piece). We think that this mismatch has limited the effect of PMT to some certain extent.
Firstly, it will hinder the model to use more surrounding information to assist the prediction of the weakened part of PR data. Secondly, it will cause that the masking of short phones is meaningless. Because the masking of these phones does not cause too much hindrance to the prediction of the modeling unit and does not encourage the model to learn more information. Moreover, these masking can not achieve the effect of PR data augmentation and even destroy the real PR data. Hence we urgently need a model structure to eliminate the granularity mismatch so that the more information produced by PMT can be considered by our model. 

Recently, many works attempt to decouple the encoder of E2E ASR and by introducing another CTC branch. Zhang {\it et al.}  \cite{decouple_pronunciation} proposes a decoupling model structure to leverage monolingual data to improve code-switching speech recogntion task. Lee {\it et al.}  \cite{Intermediate_Loss} inserts a intermediate CTC loss to regularize the model. In addition, many works try to leverage multi-modeling units to jointly optimize the E2E ASR. Lakomkin {\it et al.}  \cite{subword_regularization} point out that combining several segmentations of an utterance transcription in the loss function  to optimize the E2E ASR model may be beneficial to the model. Shubham {\it et al.} \cite{Hierarchical_CTC3} and Krishna {\it et al.}  \cite{Hierarchical_CTC2} proposes phoneme and word-piece CTC loss to joint learning based on BiLSTM model. Kubo {\it et al.} \cite{joint_phoneme_grapheme} proposes to use multi-task learning to improve generalization of the model by leveraging information from multiple labels (phoneme and grapheme). Chen {\it et al.} \cite{hybrid_modeling_units} uses a hybrid of the syllable, Chinese character, and subword as the modeling units for the end-to-end speech recognition system based on the CTC/attention multi-task learning. Nadig {\it et al.} \cite{jiont_phoneme_grapheme2} explores joint phoneme–grapheme decoding using an Encoder–Decoder network with hybrid CTC/attention mechanism.


Motivated by the above,
we adopt the hierarchical CTC based methods \cite{Hierarchical_CTC3,Hierarchical_CTC2} and propose a multi-modeling unit training (MMUT) framework fusion with PMT (PM-MMUT)
to explore more context information produced by PMT, whereby to maximize the role of PMT. In our proposed model framework (see Section \ref{sssec:mmut_framework}), we split the encoder into two parts. The first part is inserted in an intermediate phoneme-based CTC loss that allows to learn a mapping from the acoustic feature sequence to phoneme-level representation (AF-to-PLR). The second part is optimized by word-piece-based CTC and CE loss. It map the phoneme-level representation to word-piece-level representation (PLR-to-WPLR). All the losses are combined using some tunable weights. Obviously, the modeling unit of AF-to-PLR matches the masking granularity of PMT, so that the rich contextual information will be fully considered by the E2E model. 
We mainly conduct corresponding experiments on 1200 hours of Uyghur data, but also benchmark Librispeech 960 hours English task. 
The main contributions of our work can be summarized as follows: (1) We are the first to publish work on adapting the hierarchical CTC based methods \cite{Hierarchical_CTC3,Hierarchical_CTC2} to Conformer-based Encoder-Decoder E2E ASR, especially on Uyghur task. (2) We have boosted our prior work PMT, by managing to eliminate the granularity mismatch between the masking unit of PMT and the modeling unit. (3) Intensive investigations are carried out into the proposed method on both a 1200-hour Uyghur speech dataset and the 960-hour Librispeech English ASR benchmark.  The results shows a systematic improvement to verify the effectiveness of the proposed method.


\section{Model Description}
\label{sec:proposed}

\subsection{Phone Mask Training}
\label{ssec:phone_mask}

Aiming at reducing the impact of phonetic reduction in Uyghur speech, PMT randomly select a percentage of the phones and mask the corresponding speech segments in each training iteration. The masking part fill with the average value of word where the selected masked phone is located. 
The alignment information of each phone and frame is get by force alignment using HMM-DNN (nnet3) model trained by speech recognition toolkit Kaldi \cite{Kaldi}. But, the modeling unit of E2E ASR is based on word-piece, which might cause the granularity mismatch between the masking unit of PMT and modeling unit. As discuss above, this mismatch will limit the effect of PMT to some certain extent. Moreover, as for PMT, please refer to our prior work \cite{Phone_mask} for more details.

\begin{figure}[t]
\centering
 \centerline{\includegraphics[width=6.435cm]{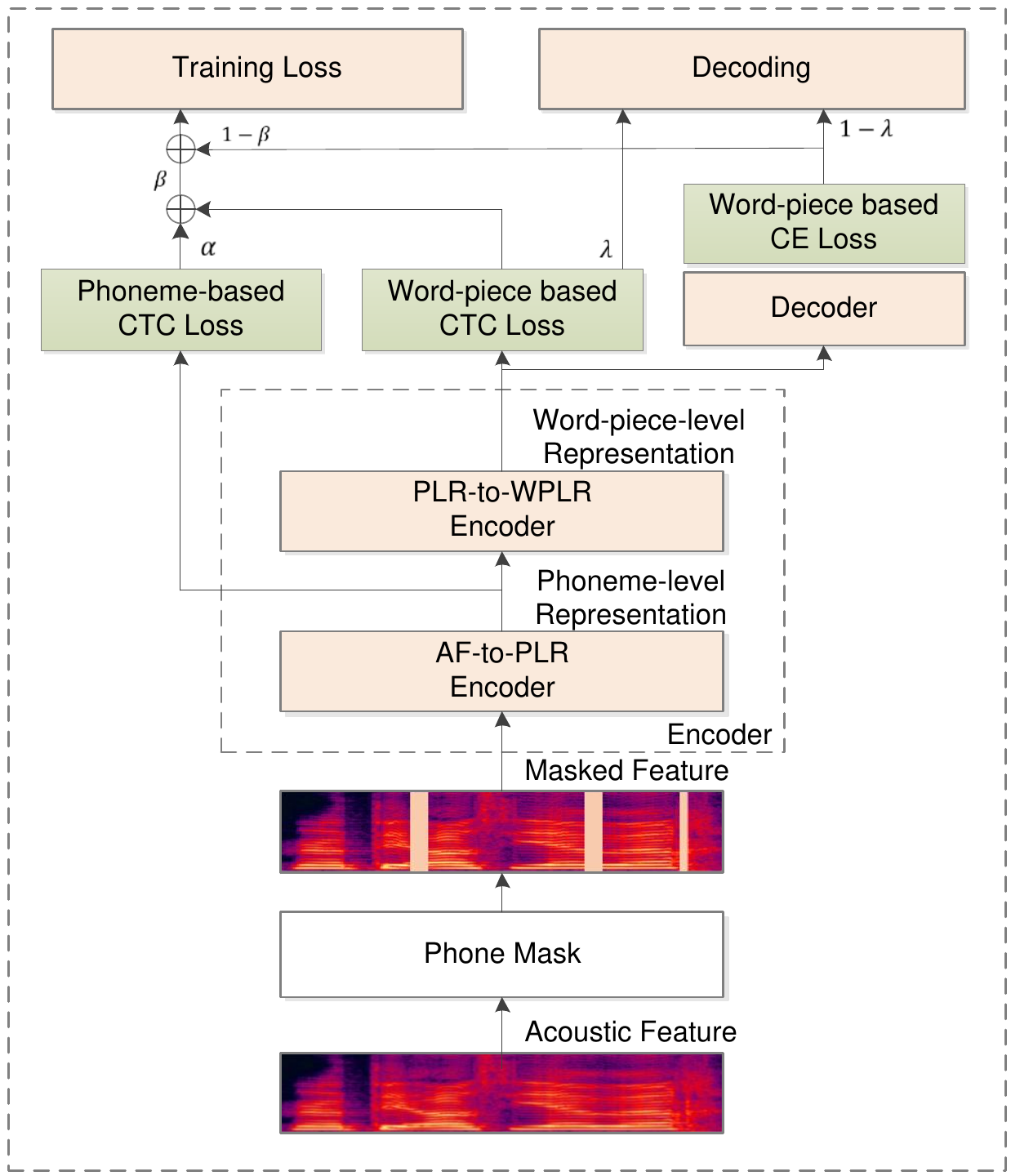}}
\caption{The structure of PM-MMUT}

\label{fig:mmut_framework}
\end{figure}

\subsection{PM-MMUT: Multi-Modeling Unit Training  Fusion with PM Training}
\label{sssec:mmut_framework}
 Figure~\ref{fig:mmut_framework} presents the structure diagram of  PM-MMUT, which is a Conformer-based E2E system. The model consists of a PM module, an encoder and a decoder. The encoder is composed by an AF-to-PLR encoder which is imposed by a phoneme-based CTC (PCTC) loss, and a PLR-to-WPLR encoder, which is imposed by a word-piece-based CTC (WPCTC) loss. The last part of the model is a standard Transformer decoder. PM-MMUT subdivides the E2E system similar to conventional HMM-DNN structure, so that the model gives full play to the role of each part, and then make the model fully explore the information contained in the data. A detailed explanation of the model is presented as follows.

In the PM module, phone mask operation is applied on the inputs to produce masked features given the input acoustic feature sequence $\mathbf X$ and the phone-level alignment information ${\mathbf{X}}_{\rm ALI}$ :
\begin{eqnarray}
    {{\mathbf {X}}}_{\rm PM} &=& {\rm PM}({\mathbf X}, {\mathbf X}_{\rm ALI}),
\end{eqnarray}
where ${\rm PM}(\cdot)$ denotes the phone mask operation and ${\mathbf X}_{\rm PM}$ represents the masked acoustic feature.

The  AF-to-PLR encoder ${\rm ENC}_{\rm A2P}(\cdot)$ learns the mapping from the acoustic feature (AF) to phoneme-level representation (PLR). It is similar to the acoustic model in traditional HMM-DNN based speech recognition system, which is made up of Conformer-based encoder in the first a few layers. It accepts ${\mathbf X}_{\rm PM}$ from the  phone mask module to produce PLR representation ${\mathbf H}_{\rm PLR}$:
\begin{eqnarray}
    \label{h_plr}
    {{\mathbf {H}}}_{\rm PLR} = {\rm ENC}_{\rm A2P}({\mathbf X}_{\rm PM}).
\end{eqnarray}
The PCTC loss is imposed here to encourage AF-to-PLR encoder 
to learn more rich phoneme-level context information brought by PMT:
\begin{eqnarray}
    \label{ctc_ph}
    {\cal {L}}_{\rm PCTC} = {\rm log} { P}_{\rm PCTC}(\rm {\cal Y}_{P}| {\mathbf H}_{\rm PLR}),
\end{eqnarray}
where ${P}_{\rm PCTC}$ considers the probability distribution over all the possible alignments based on phoneme-level label ${\cal Y}_{\rm P}$ and ${\mathbf H}_{\rm PLR}$.

The PLR-to-WPLR encoder ${\rm ENC}_{\rm P2W}(\cdot)$ is applied to learn the mapping from the phoneme-level representation (PLR) ${\mathbf H}_{\rm PLR}$ to word-piece-level representation (WPLR) ${\mathbf H}_{\rm WPLR}$, which underlying models the pronunciation lexicon in traditional HMM-DNN based ASR:
\begin{eqnarray}
    \label{h_wplr}
    {\mathbf H}_{\rm WPLR} = {\rm ENC}_{\rm P2W}({\mathbf H}_{\rm PLR}).
\end{eqnarray}
The WPCTC loss (${\cal L}_{\rm WPCTC}$) imposed on ${\mathbf H}_{\rm WPLR}$ is to regularize the model similar to general hybrid CTC/attention ASR \cite{CTC_conformer}:
\begin{eqnarray}
    \label{ctc_wp}
    {\cal L}_{\rm WPCTC} = \log P_{\rm WPCTC}({{\cal Y}_{\rm WP}}| {\mathbf H}_{\rm WPLR}),
\end{eqnarray}
where ${P}_{\rm WPCTC}$ represents the probability distribution over all possible alignments based on word-piece-level label ${\cal Y}_{\rm WP}$ and ${\mathbf H}_{\rm WPLR}$.
The total CTC loss ($\cal L_{\rm CTC}$)  combines $\cal L_{\rm PCTC}$  in Eq.(\ref{ctc_ph}) and  $\cal L_{\rm WPCTC}$ in Eq. (\ref{ctc_wp}) using a tunable trade off factor $\alpha$:
\begin{eqnarray}
  \label{ctc_loss}
    {\cal L}_{\rm CTC} = {\cal L}_{\rm WPCTC} + \alpha \times {\cal L}_{\rm PCTC},
\end{eqnarray}
where $\alpha$ is selected by hand and the  sensitive influence of $\alpha$ on PM-MMUT will be presented in the later experiments.

For the decoder, as previously mentioned,  it is a standard Transformer decoder which learns language-related information based on word-piece-level representation ${\mathbf H}_{\rm WLPR}$ and the textual information $\cal Y$. Generally, it is optimized by Cross Entropy (CE) loss:
\begin{eqnarray}
  \label{ce_loss}
    {\cal L}_{\rm CE} = \log P_{\rm DEC}({\cal Y} |{\mathbf H}_{\rm WLPR}),
\end{eqnarray}
where
\begin{eqnarray}    
    P_{\rm DEC}({\cal Y}|{\mathbf H}_{\rm WLPR}) =\prod \limits_{i=1}^L P(y_i|y_1, ..., y_{i-1}, {\mathbf H}_{\rm WLPR})
\end{eqnarray}
represents the sequence probability of  $\cal Y$ given ${\mathbf H}_{\rm WLPR}$.
The final objective is represented as a trade off between the CTC loss in Eq. (\ref{ctc_loss}) and the CE loss in Eq. (\ref{ce_loss}), as follows:
\begin{eqnarray}
    {\cal L} = \beta  \times  {\cal L}_{\rm CTC} + (1 - \beta) \times {\cal L}_{\rm CE},
\end{eqnarray}
where $\beta$ is the weight that balances
the CTC and the CE loss. In the decoding stage, only the probabilities of the decoder and WPCTC loss are combined to obtain the final output \cite{Phone_mask,recent-conformer,CTC_conformer}:
\begin{eqnarray}
    {{\hat{\cal Y}}} &= &\arg \underset{y}{\max}  \{\lambda \times \log P_{\rm DEC}(y|{\mathbf H}_{\rm WLPR}) \\
    &&+ (1-\lambda)\times\log P_{ \rm WPCTC}(y|{\mathbf H}_{\rm WLPR})\},
\end{eqnarray}
where  $\lambda$ is a tunable hyper-parameter controlling score balance between CTC and attention. Following \cite{Phone_mask,recent-conformer}, $\lambda$ is set to 0.6 over all experiments.

\section{Experiments and Results}
\label{sec:exp_and_result}

\subsection{Data Description}
\label{sec:data_description}

The proposed method is evaluated on both Uyghur large vocabulary speech recognition and the standard Librispeech speech recognition benchmark \cite{Librispeech-paper}. For Uyghur speech recognition experiments, we use the same Uyghur speech corpus in our previous work \cite{Phone_mask}. The database contains speech sampled at 16k Hz with a duration of 1 200 hours. And the corpus contains 1 198 582 utterances read by 65 089 speakers. As for the evaluation set, we use reading speech (Read-Test), spontaneous speech (Oral-Test) and THUYG-20 \cite{Thuyg-20} test (THUYG-Test) similar to our previous configurations \cite{Phone_mask}. For English ASR evaluation, we experiment on the Librispeech 960-hour benchmark  and test on test-clean/other and dev-clean/other. The details of the datasets are shown in Table~\ref{data_description}.
\begin{table}[t]
  \caption{The datasets }
  \renewcommand\tabcolsep{2.0pt}
  \centerline{}
  \label{data_description}
  \centering
  \begin{tabular}{|l|l|cc|}
    \hline
    &{{Data Type}}  & {{Dur (Hrs)}} & {{Domain}} \\
    \hline
    \hline
    &Train & 1200 & Read/Clean \\
   Uyghur &Read-Test & 3& Read/Clean \\
    &Oral-Test & 5& Oral/Noisy \\
    \hline
    \hline
    &Train-960 & 961 & Read/Clear\&Noisy \\
    English &Test/Dev-clean & 5.4 & Read/Clear \\
    &Test-other  & 5.1 & Read/Noisy \\
    &Dev-other  & 5.3 & Read/Noisy\\
    \hline
  \end{tabular}
\end{table}

\subsection{Experiment setup}
\label{ssec:experiment_setup}
For Uyghur speech recognition task, following our previous setups \cite{Phone_mask}, the experiments use 40 Mel Frequency Cepstral Coefficients (MFCCs) over 25 ms frames with 10 ms stride to each of which cepstral mean and variance normalization (CMVN) is applied. In English tasks, following \cite{recent-conformer}, we use  80-dimensional logmel spectral energies plus 3 extra features for pitch information as acoustic features input. Following \cite{Phone_mask,recent-conformer}, the trade off weight $\beta$ is set to 0.3 over all the tasks. For the E2E configuration, we use a similar setup in our work \cite{Phone_mask} in the Uyghur ASR experiment, and follow the ESPnet official configuration \cite{recent-conformer} for the Librispeech task.
All the E2E models (Encoders = 12, Decoders = 6, Aheads = 8, $d^{att}$= 512) are trained by using  ESPnet1 \cite{espnet} on 4 P40 GPUs for the Uyghur task and 8 M40 GPUs for the English task. No external language models are used. Then the pre-trained Conformer model \footnote{https://github.com/espnet/espnet/tree/master/egs/librispeech/asr1} from the latest ESPnet official version, is used as an important baseline. The output tokens are 5000 BPE tokens produced by unigram model using sentencepiece \cite{SentencePiece} in all tasks. In addition, we use $N_{\rm A2P}$ to represent the number of the layers of AF-to-PLR encoder and the number of PLR-to-WPLR encoder is 12 - $N_{\rm A2P}$. Especially, both AF-to-PLR and PLR-to-WPLR encoder share the 12 Conformer encoder layers if the $N_{\rm A2P}$ equal to 12.

\subsection{Results on Uyghur ASR}
\label{ssec:result_12k_ug}

\subsubsection{Base experiments}
\label{sssec:Baseline}

We experiment with the basic MMUT, PM-MMUT and compare the results with the baselines from our prior work \cite{Phone_mask}. In Table~\ref{phone:base_exp}, it can be seen that the basic MMUT, AF-to-PLR alleviates granularity mismatch between the reduction unit (phoneme) and the modeling unit (word-piece), achieves  obvious improvements over the baselines. However, it demonstrates inferior performance gain than pure PMT which yield the best results in \cite{Phone_mask}. But, when we combine PMT with the basic MMUT, the basic PM-MMUT performs better than both the basic MMUT and PMT. The results suggest that MMUT and PMT are complementary, which is consistent with our expectation. 

\begin{table}[t]
	\renewcommand\arraystretch{0.9}	
  \caption{WERs from the basic MMUT and PM-MMUT}
  \renewcommand\tabcolsep{2.0pt}
  \label{phone:base_exp}
  \centerline{}
  \centering
  \begin{tabular}{ |p{95pt}  p{40pt}|p{37pt} p{37pt}|}
  \hline
    {SYSTEM} &{\hfil $(\alpha,N_{\rm A2P})$} & {\hfil Read-Test}  & {\hfil Oral-Test}   \\
    \hline
    \hline
    Chain         &\hfil           &\hfil 26.0           &\hfil 47.9           \\
    Conformer (Grapheme)  &\hfil      & \hfil  26.8            &\hfil 45.0            \\
    \hline
    \hline
    Conformer (Word-piece)  &\hfil    &\hfil  25.4            & \hfil 44.1            \\
    \quad + SpecAugment  &\hfil            &\hfil  25.2            &\hfil 41.6            \\
    \quad + PMT         &\hfil               &\hfil  24.0            &\hfil  38.4            \\
    \quad + MMUT  & \hfil (0.5,12)              & \hfil 24.9            &\hfil 41.6            \\
    \quad + PM-MMUT & \hfil (0.5,12)         & \textbf{\hfil 23.9}   &\textbf{\hfil 37.5}   \\
    \hline
  \end{tabular}
\vspace{0.3cm}
  \caption{WERs by varing  $N_{\rm A2P}$}
\vspace{-0.3cm}
  \renewcommand\tabcolsep{2.0pt}
  \label{phone:af-to-plr-layer}
  \centerline{}
  \centering
  \begin{tabular}{ |p{95pt} p{40pt}  |p{37pt} p{37pt}|}
  \hline
    {SYSTEM} &{ $(\alpha,N_{\rm A2P})$}  & {\hfil Read-Test}  & {\hfil Oral-Test}   \\
    \hline
    \hline
    Conformer (Word-piece)  & \hfil     &\hfil  25.4            & \hfil 44.1            \\
    \quad + PMT  & \hfil                    &\hfil  24.0            &\hfil  38.4            \\
    \quad + PM-MMUT &  (0.5, 12)        &\hfil 23.9   &\hfil 37.5 \\
    \quad + PM-MMUT &  (0.5, ~~3)          &\hfil 24.1   &\hfil 39.7 \\
    \quad + PM-MMUT &  (0.5, ~~6)          &\hfil 23.9   &\hfil 37.5 \\
    \quad + PM-MMUT &  (0.5, ~~8)          &\hfil 24.0   &\hfil 37.1 \\
    \quad + PM-MMUT &  (0.5, 10)         &\textbf{\hfil 23.7}   &\textbf{\hfil 36.8} \\
    \quad + PM-MMUT &  (0.5, 11)         &\hfil 24.4   &\hfil 37.2 \\
    \hline
  \end{tabular}
\vspace{-0.2cm}
\end{table}

\subsubsection{Results on subdivision of the encoder, $\alpha$ and granularity}
\label{sssec:PM_MMUT}

The subdivision of the encoder, i.e. the number of $N_{\rm A2P}$, and weight $\alpha$ directly affects phoneme-level representation, and consequently, they indirectly influence the PLR-to-WPLR encoder and the decoder.

We experiment with different $N_{\rm A2P}$=3, 6, 8, 10, 11, 12. The number of layers is closely related to the learning ability of AF-to-PLR and PLR-to-WPLR encoder. Larger $N_{\rm A2P}$  generally means  stronger representation learning capability of AF-to-PLR encoder. But this might weaken the PLR-to-WPLR encoder. Therefore, we need a balance in deciding the number of layers in AF-to-PLR and PLR-to-WPLR encoder. As shown in Table~\ref{phone:af-to-plr-layer},  $N_{\rm A2P}$=10 shows the best result, which is consistent with our expectation: AF-to-PLR encoder needs a strong representation learning capability  in mapping from the acoustic feature to phoneme-level representation, and PLR-to-WPLR encoder is to map phoneme-level representation to word-piece-level representation, which is a relatively easy task.

\begin{table}[t]
  \caption{WERs from PM-MMUT, intermediate CTC loss, and word-piece-CTC (PM-MMUT-WP)}
  \renewcommand\tabcolsep{2.0pt}
  \label{phone:pm_mmt-inter_ctc}
  \centerline{}
  \centering
  \begin{tabular}{ |p{95pt}  p{40pt} |p{37pt} p{37pt}|}
  \hline
    {\hfil SYSTEM} &{\hfil $(\alpha,N_{\rm A2P})$} & {\hfil Read-Test}  & {\hfil Oral-Test}   \\
    \hline
    \hline
    Conformer (Word-piece)   &\hfil  &\hfil      25.4            & \hfil 44.1            \\
    \quad + PMT  &\hfil    &\hfil  24.0            & \hfil 38.4            \\
    \quad\quad+ Inter. CTC \cite{Intermediate_Loss}  &\hfil        &\hfil 23.7   &\hfil 38.0 \\
    \quad + PM-MMUT    &\hfil (0.5,10)         &\textbf{\hfil 23.7}   &\textbf{\hfil 36.8} \\
    \quad + PM-MMUT-WP &\hfil (0.5,10)      &\hfil 23.9   &\hfil 38.3 \\
    \hline
  \end{tabular}
\vspace{0.3cm}
  \caption{WERs by comparing with speed perturbation}
\vspace{-0.3cm}
  \renewcommand\tabcolsep{2.0pt}
  \label{phone:speed_perturbation}
  \centering
  \begin{tabular}{ |p{95pt}  p{40pt}  |p{37pt} p{37pt}|}
  \hline
    {SYSTEM} &{\hfil $(\alpha,N_{\rm A2P})$} & {\hfil Read-Test}  & {\hfil Oral-Test}   \\
    \hline
    \hline
    Conformer (Word-piece)  &\hfil    &\hfil  25.4            & \hfil 44.1            \\
    \quad Speed Perturbation  &\hfil   &\hfil  24.5            & \hfil 39.9  \\
    \quad\quad + PMT      &\hfil            &\hfil  23.5            & \hfil 37.0 \\
    \quad\quad + PM-MMUT &\hfil (0.5,10)  &\textbf{\hfil 23.3}   &\textbf{\hfil 35.9} \\
    \hline
  \end{tabular}

\vspace{-0.5cm}  
\end{table}

\begin{figure}[htp]
\centering
  \centerline{\includegraphics[width=8.5cm]{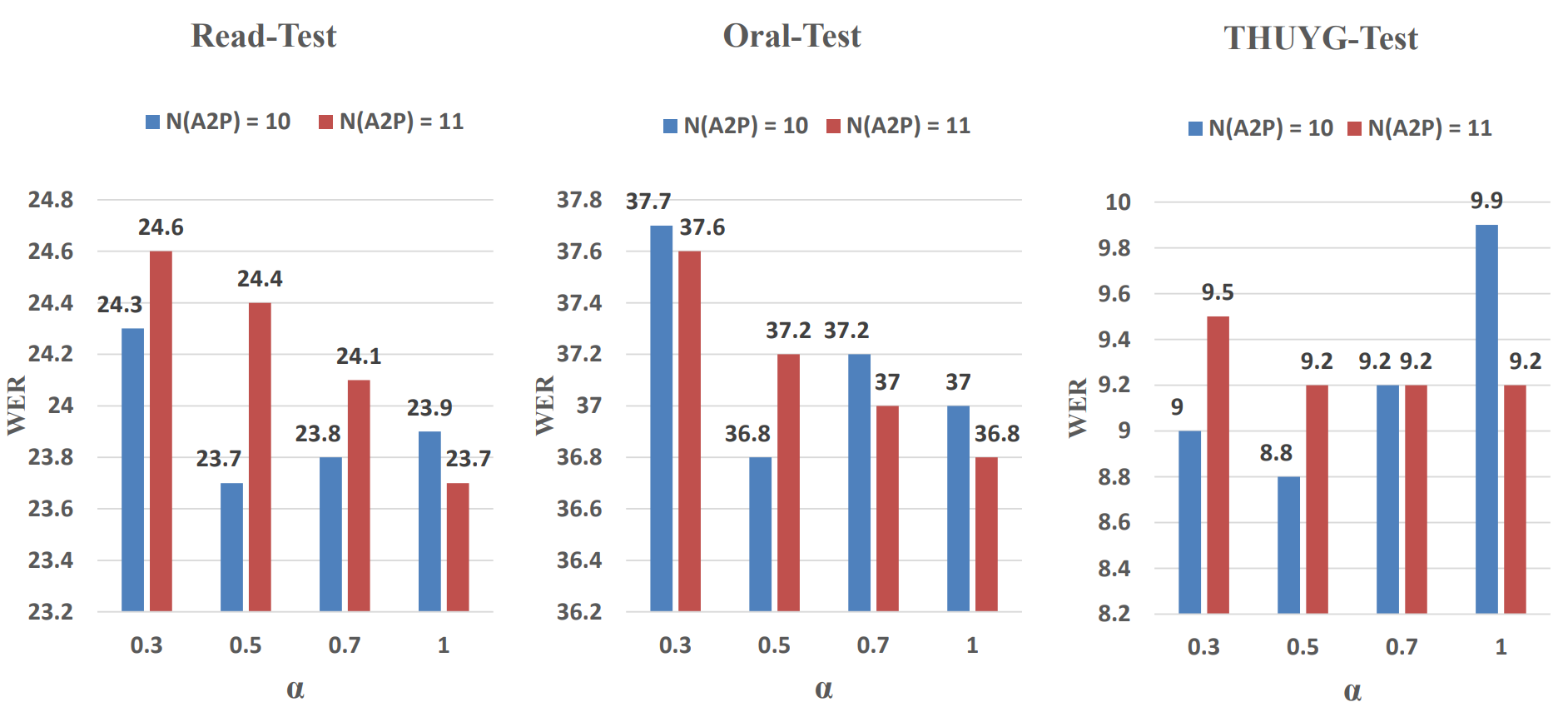}}
\caption{Results with different  $\alpha$ when ${ N}_{\rm A2P}$=10 and 11 (we display the ${ N}_{\rm A2P}$ with N(A2P) in this figure).}
\label{fig:res}

\vspace{-0.3cm}
\end{figure}

As for $\alpha$, we fix $N_{\rm A2P}$ to 10 and 11. Then, we conduct experiments with  $\alpha$ varying from 0.3, 0.5 and 0.7 to 1.0. The results are shown in Figure~\ref{fig:res}. When $N_{\rm A2P}$ is set to 10, the performance becomes worse with the increasing of $\alpha$. This might be because the two layers of PLR-to-WPLR encoder is still stronger. If $\alpha$ increases,  the AF-to-PLR encoder  pays more attention to the learning of phoneme-level representation, which is more likely to make the PLR-to-WPLR encoder over-fitting. When $N_{\rm A2P}$ is set to 11, which might weaken PLR-to-WPLR encoder because there is only one layer of Conformer encoder, thus the degraded result is obtained. This further justifies our above analysis. The results is also consistent with our hypothesis that the balance between AF-to-PLR and PLR-to-WPLR encoder is important in the proposed PM-MMUT framework.

We further carried out an experimental comparison of intermediate CTC (Inter. CTC) \cite{Intermediate_Loss}, the proposed PM-MMUT, and PM-MMUT-WP (replacing the phoneme-based CTC of PM-MMUT with word-piece-based CTC). The intermediate CTC inserts an CTC loss to regularize the model.
As shown in Table~\ref{phone:pm_mmt-inter_ctc}, the intermediate CTC  can improve on PMT, but is still worse than PM-MMUT. When we replace the phoneme-CTC with word-piece-CTC, there appears to be very little help on PMT. This further verified that PM-MMUT alleviates the granularity mismatch between masking unit of PMT and modeling unit. Then the proposed PM-MMUT makes the modeling unit more compatible with  PMT to achieve improved performance.


\subsubsection{Results on PM-MMUT with speed perturbation}
\label{sssec:exp_sp}

As we have mentioned in our prior work \cite{Phone_mask}, it is naturally to consider the utilization of the standard speed perturbation \cite{speed-perturb} to augment the speech data of variant speed. In \cite{Phone_mask}, we demonstrate that PMT and speed perturbation are complimentary. Here we also experiment on PM-MMUT after speed perturbation.  We can see in Table~\ref{phone:speed_perturbation} that a significant improvement using PM-MMUT comparing with PM after speed perturbation.


\subsection{Results on English ASR }
\label{ssec:result_libri}

To further demonstrate the effectiveness of the proposed method, we present the results on a popular ASR benchmark Librispeech 960 hours task. Table~\ref{phone:libri} shows the results. We can see that PMT is still useful in Librispeech due to phonetic reduction phenomenon exists in English yet, but the performance of PMT is worse than SpecAugment. However, when we fuse PMT and MMUT, the overall result of PM-MMUT becomes better than SpecAugment. This is consistent with our motivation that the PM-MMUT alleviates the granularity mismatch between the masking unit of PMT and the modeling unit, which will encourage the model to explore more context information produced by PMT.
And we also can see that PMT, PM-MMUT are both complementary with SpecAugment. Finally, our  model achieves the best performance without external language model, which achieves about 10\% relative WER reduction on all the tests comparing with the official ESPnet1 pre-trained model.
\begin{table}[t]
  \caption{WERs on Librispeech 960 with pure PMT and PM-MMUT}
  \renewcommand\tabcolsep{2.0pt}
  \label{phone:libri}
  \centerline{}
  \centering
  \begin{tabular}{ |p{84pt} p{40pt} |p{20pt} p{20pt} |p{20pt} p{22pt}|}
  \hline
      & & \multicolumn{2}{c|}{Dev}  & \multicolumn{2}{c|}{Test}   \\
    \hline
    {SYSTEM} & {\hfil $(\alpha,N_{\rm A2P})$}  & {\hfil clean} &{\hfil other}  & {\hfil clean} &{\hfil other}   \\
    \hline
    \hline
    Large Transformer \cite{semantic_paper} & \hfil      & {\hfil 2.64}     & {\hfil 6.90}          & {\hfil 2.74}          & {\hfil 6.65}   \\
    ESPnet model &\hfil     & {\hfil 2.6}  & {\hfil 6.8}      & {\hfil 2.8}           & {\hfil 7.0}    \\
    Our base             &\hfil          & {\hfil 3.3}  & {\hfil 9.7}      & {\hfil 3.5}           & {\hfil 9.6}    \\
    \quad + SpecAugment &\hfil           & {\hfil 2.7}  & {\hfil 6.9}      & {\hfil 2.9}           & {\hfil 7.0}    \\
    \quad + PMT        &\hfil         & {\hfil 2.8}  & {\hfil 7.5}      & {\hfil 3.0}           & {\hfil 7.7}    \\
    \quad\quad + SpecAugment &\hfil     & {\hfil 2.4}  & {\hfil 6.3}      & {\hfil 2.8}           & {\hfil \textbf{6.2}}  \\
    \quad + PM-MMUT      &\hfil (0.5,10)  &{\hfil 2.6} & {\hfil 6.5}   &{\hfil 2.9}       & {\hfil 6.7}    \\
    \quad\quad + SpecAugment &\hfil     & {\hfil \textbf{2.3}} & {\hfil \textbf{6.1}}  
          & {\hfil \textbf{2.6}}  & {\hfil \textbf{6.2}}  \\
    \hline
  \end{tabular}
 
\vspace{-0.5cm} 
  
\end{table}
\section{Conclusions}
\label{sec:conclusion}

In this paper, aiming at boosting our prior work phone mask training (PMT), we propose an effective multi-modeling unit learning architecture fusion with PMT (PM-MMUT) for E2E speech recognition. During training, we will split Encoder into two parts including acoustic feature sequences to phoneme-level representation (AF-to-PLR) and phoneme-level representation to word-piece-level representation (PLR-to-WPLR). Therefore, it allows AF-to-PLR to be enforced by an intermediate phoneme-based CTC loss 
to learn the rich phoneme-level contextual information brought by PMT. We have carried out Uyghur ASR experiments on both reading and spontaneous speech. Results show that the proposed method improves the performance of Uyghur ASR obviously, especially on the spontaneous speech. Extensive investigations into ASR benchmark Librispeech 960 hours have also been carried out and confirm the effectiveness of the proposed method.  According to the analysis, the PM-MMUT is helpful in improving the phonetic-reduction-robustness of Uyghur and English E2E ASR. 

\section{Acknowledgements}

This work was supported by the National Key R\&D Program of China (2020AAA0107902), Opening Project of Key Laboratory of Xinjiang, China (2020D04047), and Natural Science Foundation of China (61663044, 61761041).


\bibliographystyle{IEEEtran}

\bibliography{mybib}


\end{document}